\documentclass[aps,prl,reprint,groupedaddress,floatfix]{revtex4-2}

\usepackage{amsmath}
\usepackage{amssymb}
\usepackage{bm}
\usepackage{graphicx}
\usepackage{xcolor}

\newcommand{\rmm}{\mathrm{m}}                    
\newcommand{\taum}{\tau_{\rmm}}
\newcommand{\bq}{\bm{q}}
\newcommand{\br}{\bm{r}}
\newcommand{\bD}{\bm{D}}
\newcommand{\bQ}{\bm{Q}}
\newcommand{\bd}{\bm{d}}
\newcommand{\bOmega}{\boldsymbol{\Omega}}
\newcommand{\bmeta}{\boldsymbol{\eta}}

\newcommand{\traj}{\mathrm{traj}}
\newcommand{\opt}{\mathrm{opt}}

\begin{document}

\title{Optical-Memory Transport Imaging: Extension to Stochastic Diffusion}

\author{Haichun Liu}
\email{haichun@kth.se}
\author{Jerker Widengren}
\affiliation{Experimental Biomolecular Physics, Department of Applied Physics, KTH Royal Institute of Technology, S-106 91 Stockholm, Sweden}

\begin{abstract}
Finite-memory optical tracers can encode transport properties through their previously experienced excitation. Using such memory features, harboured in the population of excitable long-lived electronic states within the emitters, instantaneous localization assessments are not needed to determine their transport properties. Here we formulate this principle for monitoring of stochastic transport extracted from transport-history integral over conditional transport histories. Under structured illumination, this history yields a measurable transfer-function response from which scalar and tensor diffusion are reconstructed without time-resolved acquisition. Fisher-information analysis identifies the optimal operating regime and yields first-principles, parameter-free predictions of scalar and tensor reconstruction precision, in quantitative agreement with independent Monte Carlo simulations. These results establish finite optical memory as a general principle for stochastic transport imaging.
\end{abstract}

\maketitle

\section*{Introduction}

The central challenge of diffusion imaging is to convert stochastic transport into measurable observables. Existing optical approaches achieve this by introducing an explicit temporal coordinate. Fourier-transform fluorescence recovery after patterned photobleaching (FT-FRAP)~\cite{FTFRAP} measures recovery from a written bleach pattern, and its light-sheet extension~\cite{LiFTFRAP} further reconstructs three-dimensional diffusion tensors from a volumetric bleach-recovery sequence. Fluorescence correlation spectroscopy (FCS)~\cite{FCS} extracts transport from the temporal autocorrelation of fluorescence fluctuations within an observation volume, while raster image correlation spectroscopy (RICS)~\cite{RICS} extends correlation analysis to the spatiotemporal sampling imposed by confocal raster scanning. Differential dynamic microscopy (DDM)~\cite{DDM} extracts transport from temporal image correlations, while forced Rayleigh scattering~\cite{FRS} and transient-grating spectroscopy~\cite{TG} analyze the temporal decay of optically generated gratings. Although implemented differently, these techniques share a common measurement requirement: transport is inferred from an externally resolved temporal evolution of the measured signal rather than from excitation history retained in the tracer's optical state.

Optical memory has previously been exploited for molecular mobility measurements using long-lived photoinduced states. In transient-state (TRAST) measurements, molecular passages of fluorescence emitters through a localized excitation volume can modulate their reversible photo-induced dark-state populations. This allows the mobility of the emitters to be inferred from time-averaged fluorescence and through the excitation history retained in their photophysical states~\cite{TRASTdiffusion}.

Building on the broader principle of transport-history encoding, we recently introduced optical-memory transport (OMT) imaging~\cite{OMTflow}, in which finite optical memory replaces externally resolved time by integrating excitation accumulated along transport trajectories through the intrinsic photophysics of the tracer. Under structured illumination, deterministic transport histories are thereby converted into frequency-domain optical measurables, enabling quantitative flow measurements from steady-state images.

The key simplification in deterministic transport is that the tracer motion is characterized by a single upstream transport history, which can be directly encoded by the tracer's finite optical memory. Stochastic transport fundamentally changes this picture. Instead of a single upstream transport history, transport is characterized by an ensemble of possible histories. The central question therefore becomes whether finite optical memory can perform the required temporal integration over this ensemble, thereby replacing the explicit temporal sampling on which existing diffusion-imaging methods rely. Answering this question requires extending OMT from a single deterministic transport history to an ensemble description of transport history.

Here we establish the stochastic theory of OMT imaging through a transport-history integral that generalizes deterministic trajectories to stochastic transport ensembles. Under spatially structured illumination, stochastic transport is encoded in the spatial relationship between the excitation field and the resulting optical signal, enabling quantitative reconstruction of scalar and tensor diffusion through calibrated inversion. Fisher-information analysis further identifies optimal measurement conditions and yields first-principles, parameter-free predictions of reconstruction precision that quantitatively reproduce independent Monte Carlo results. Together, these results establish the general measurement principles for stochastic transport imaging using the finite optical memory of moving tracers, whereby recent excitation history, i.e., transport relative to an excitation field pattern, is retained in their optical states. For the emitters considered here, this is encoded in long-lived excited-state populations over timescales set by the lifetimes of these states.

\section{OMT Formulation for Stochastic Transport}
\label{sec:formulation}

\begin{figure}[t]
\centering
\includegraphics[width=1.0\columnwidth]{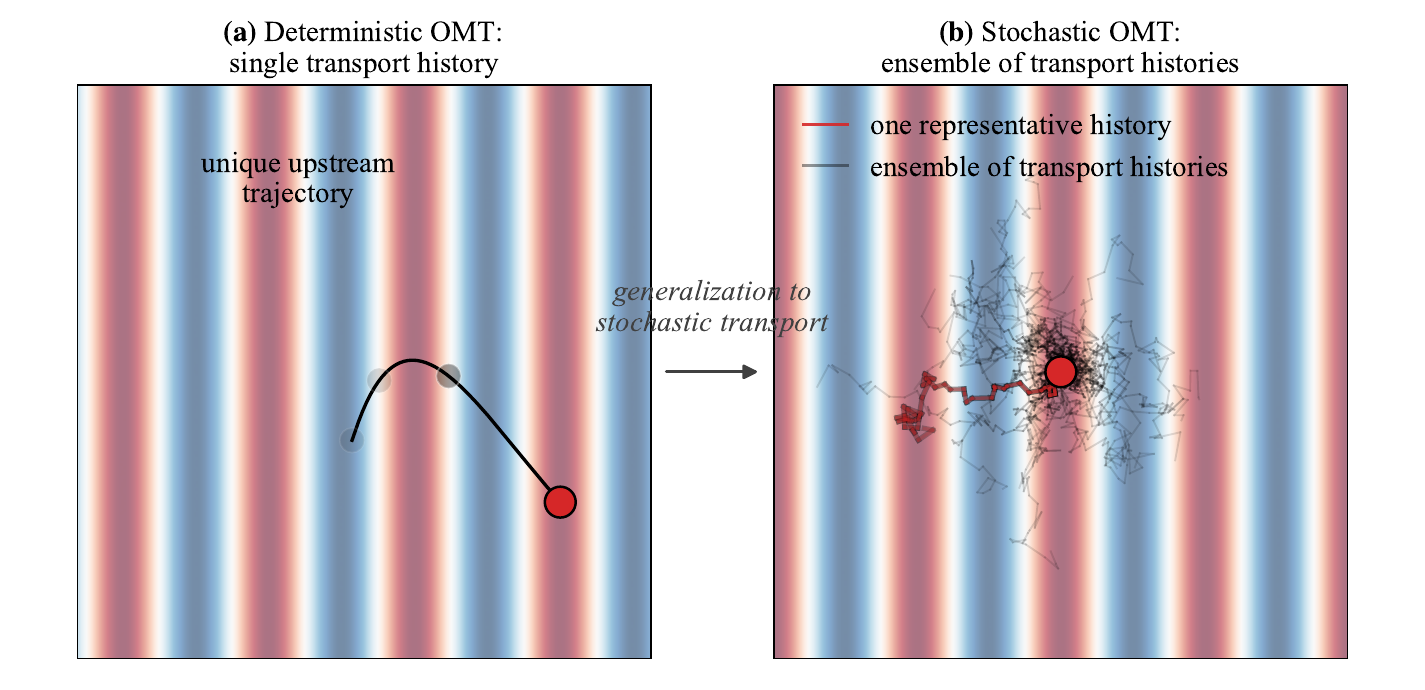}
\caption{Transport-history formulation. (a) Deterministic optical-memory transport (OMT), in which finite optical memory integrates excitation along a unique upstream transport trajectory~\cite{OMTflow}. (b) Stochastic OMT, in which the unique trajectory is replaced by an ensemble of upstream transport histories described by the conditional transport propagator.}
\label{fig:1}
\end{figure}

Finite optical memory causes tracer responses to depend on transport history rather than instantaneous local transport. For deterministic transport, this history follows a unique upstream trajectory (Fig.~\ref{fig:1}(a)), leading to the original OMT relation~\cite{OMTflow},
\begin{equation}
\delta S(\br,t) = \int_0^\infty h(\tau)\,\delta I[\br_{\traj}(t-\tau),\,t-\tau]\,d\tau,
\label{eq:1}
\end{equation}
where $\delta S(\br,t)$ is the memory-dependent optical signal modulation from a tracer located at $\br$ at time $t$, $\br_{\traj}(t-\tau)$ denotes its position at the earlier time $t-\tau$ along a trajectory terminating at $\br$ at time $t$, $\delta I(\br,t)$ is the excitation modulation about a background level,
 $h(\tau)$ is the effective impulse-response kernel of the optical tracer, and $\tau$ is the time delay.

For stochastic transport, the upstream trajectory is no longer unique. Instead, transport is described by an ensemble of possible upstream histories consistent with the same observation (Fig.~\ref{fig:1}(b)). The required generalization replaces the deterministic trajectory by the conditional transport propagator, $P_{\mathrm{b}}(\br',t-\tau\mid\br,t)$, giving the probability density that a tracer detected at position $\br$ at time $t$ occupied position $\br'$ a time $\tau$ earlier~\cite{VanKampen}. Substituting into Eq.~\eqref{eq:1} yields
\begin{equation}
\begin{split}
\delta S(\br,t) = \int_0^\infty h(\tau)\Big[&\int P_{\mathrm{b}}(\br',t-\tau\mid\br,t)\\
&\times\,\delta I(\br',t-\tau)\,d\br'\Big]d\tau.
\end{split}
\label{eq:2}
\end{equation}
Equation~\eqref{eq:2} defines the transport-history integral. Instead of integrating excitation along a single upstream trajectory, it averages excitation over all transport histories consistent with the detected tracer. As a consistency check, the deterministic theory is recovered immediately when $P_{\mathrm{b}}(\br',t-\tau\mid\br,t)=\delta[\br'-\br_{\traj}(t-\tau)]$, for which Eq.~\eqref{eq:2} reduces identically to Eq.~\eqref{eq:1}. Deterministic transport therefore is the zero-uncertainty limit of the transport-history integral.

Under structured illumination, the excitation decomposes into independent spatial Fourier components, which, by linearity, can be analyzed independently. We therefore consider a single Fourier mode of arbitrary orientation  
\begin{equation*}
\delta I(\br,t) = \operatorname{Re}\{I_{\bq}(t)\,e^{i\bq\cdot\br}\},
\end{equation*}
where $\bq$ is the wavevector of the illumination pattern applied onto the sample. Substituting this Fourier component into Eq.~\eqref{eq:2} and introducing the displacement coordinate, $\br'=\br+\Delta\br$, separates the local spatial phase from the stochastic displacement statistics (see Sec.~S1 of the Supplemental Material), giving
\begin{equation}
\delta S(\br,t) = \operatorname{Re}\left\{e^{i\bq\cdot\br}\int_0^\infty h(\tau)\,I_{\bq}(t-\tau)\,\Phi_{\mathrm{b}}(\bq,\tau;\br,t)\,d\tau\right\}.
\label{eq:3}
\end{equation}
Here, $\Phi_{\mathrm{b}}(\bq,\tau;\br,t)=\langle e^{i\bq\cdot\Delta\br(\tau)}\rangle_{\mathrm{b}}$ is the conditional displacement characteristic function, equivalently the spatial Fourier transform of the backward propagator,
\begin{equation}
\Phi_{\mathrm{b}}(\bq,\tau;\br,t) = \int P_{\mathrm{b}}(\br+\Delta\br,t-\tau\mid\br,t)\,e^{i\bq\cdot\Delta\br}\,d\Delta\br,
\label{eq:4}
\end{equation}
with the average conditioned on detection at $(\br,t)$.

For stationary structured illumination, $I_{\bq}(t)=I_{\bq}$, the detected signal retains the imposed spatial frequency,
\begin{equation*}
\delta S(\br,t) = \operatorname{Re}\{I_{\bq}\,R(\bq;\br,t)\,e^{i\bq\cdot\br}\},
\end{equation*}
with
\begin{equation}
R(\bq;\br,t) = \int_0^\infty h(\tau)\,\Phi_{\mathrm{b}}(\bq,\tau;\br,t)\,d\tau.
\label{eq:5}
\end{equation}
Equation~\eqref{eq:5} defines the stochastic OMT response under structured illumination: the measured Fourier component is determined by the optical-memory weighting $h(\tau)$ of the conditional displacement characteristic function $\Phi_{\mathrm{b}}(\bq,\tau;\br,t)$.

\section{OMT Forward Model for Stochastic Diffusion}
\label{sec:forward}

\subsection{Isotropic Brownian diffusion}
\label{sec:isotropic}

For homogeneous transport with stationary displacement statistics, the conditional displacement characteristic function becomes independent of the detection position and time, and Eq.~\eqref{eq:5} reduces to
\begin{equation*}
R(q) = \int_0^\infty h(\tau)\,\Phi(\bq,\tau)\,d\tau.
\end{equation*}
For isotropic Brownian diffusion with diffusion coefficient $D$, the Gaussian displacement propagator~\cite{VanKampen} (Supplemental Material Sec.~S2) has characteristic function
\begin{equation*}
\Phi(\bq,\tau) = \exp(-Dq^2\tau),
\end{equation*}
so that Brownian diffusion enters the OMT response only through the transport variable $T=Dq^2$, via the optical-memory transfer function
\begin{equation*}
H(T) = \int_0^\infty h(\tau)e^{-T\tau}\,d\tau,
\end{equation*}
giving
\begin{equation}
R(\bq) = H(Dq^2) = H(T).
\label{eq:6}
\end{equation}
Equation~\eqref{eq:6} establishes the general forward model for homogeneous isotropic Brownian diffusion: transport determines the scalar variable $T$, whereas tracer photophysics determines the transfer function $H(T)$.

For the single-exponential reference kernel with memory time $\taum$,
\begin{equation*}
h(\tau) = \frac{1}{\taum}e^{-\tau/\taum},
\end{equation*}
it is convenient to introduce the dimensionless transport-memory parameter
\begin{equation*}
\Omega = Dq^2\taum = T\taum.
\end{equation*}
The response then becomes
\begin{equation}
R(\Omega) = \frac{1}{1+\Omega}.
\label{eq:7}
\end{equation}
Equation~\eqref{eq:7} makes several direct physical predictions. In the absence of diffusion, $D=0$, and hence $\Omega=0$, giving $R=1$: the imposed spatial modulation is preserved because tracers do not sample different illumination phases during the optical-memory time. For finite diffusion, $D>0$, Brownian motion causes tracers to sample an increasing range of illumination phases before emission, producing ensemble averaging of the spatial modulation and therefore $R<1$. The attenuation increases monotonically with $D$, $q^2$, and $\taum$, and the modulation vanishes asymptotically as $\Omega\to\infty$.

Because Brownian displacements are symmetric, this averaging attenuates the modulation amplitude without producing a systematic phase shift. Diffusion is therefore encoded directly in the contrast reduction relative to the zero-diffusion response.

To test these predictions, we performed Monte Carlo simulations in which Brownian trajectories with prescribed diffusivity were propagated under sinusoidal structured illumination, and the detected emission was generated by weighting the excitation history along each trajectory with the single-exponential memory kernel (Fig.~\ref{fig:2}(a), Sec.~S6 of the Supplemental Material). Representative emission images in Figs.~\ref{fig:2}(b) and \ref{fig:2}(c) show the predicted contrast attenuation from weak to strong transport-memory coupling. For quantitative comparison, the modulation amplitude extracted from each simulated image was normalized to its zero-diffusion value to obtain $R$. Figure~\ref{fig:2}(d) shows quantitative agreement between these Monte Carlo responses and Eq.~\eqref{eq:7} over the investigated range of $\Omega$.

\begin{figure}[t]
\centering
\includegraphics[width=1\columnwidth]{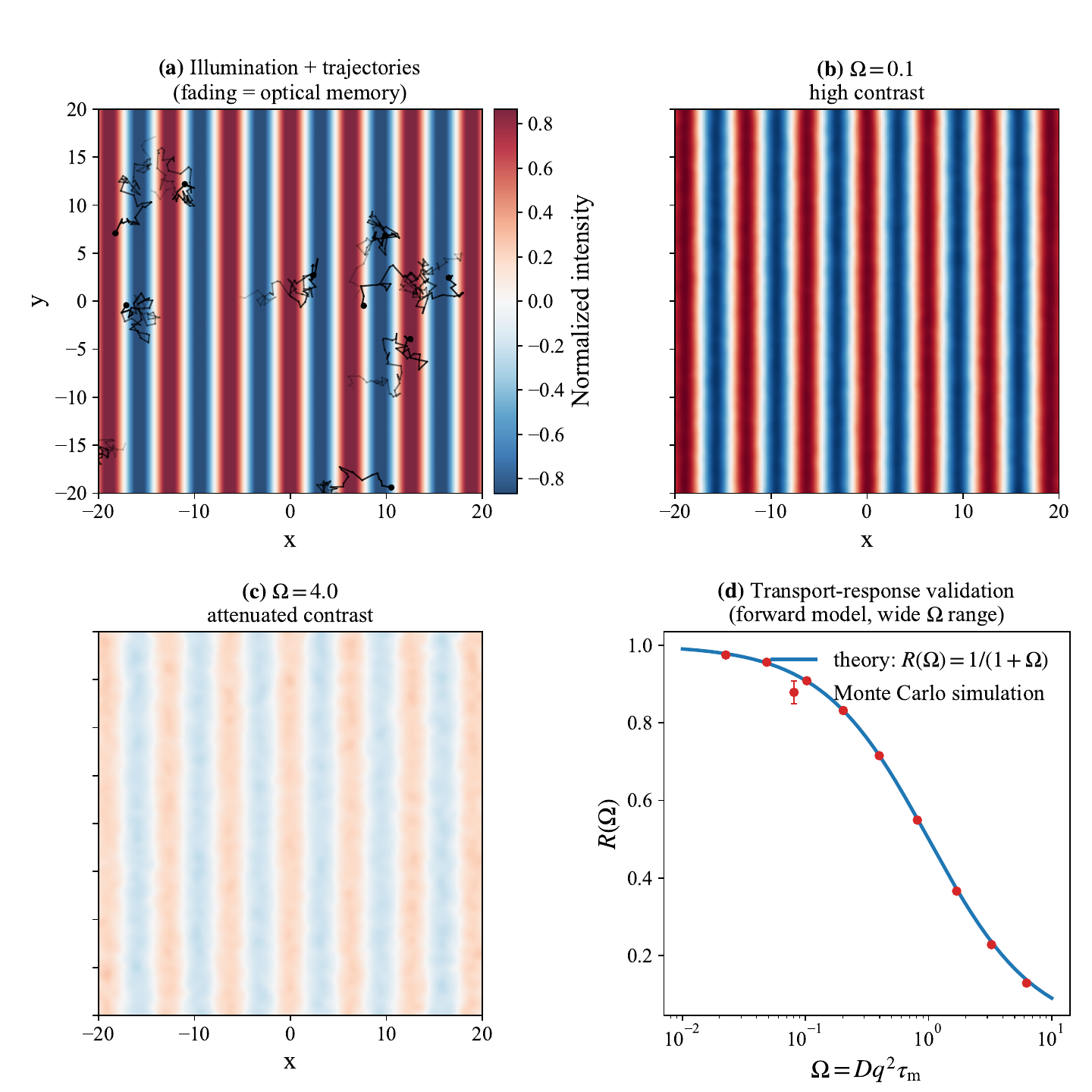}
\caption{Forward model for isotropic Brownian diffusion. (a) Monte Carlo simulation scheme for Brownian diffusion under structured illumination. (b,c) Representative emission images for weak and strong transport-memory coupling, illustrating progressive attenuation of modulation contrast with increasing transport-memory parameter $\Omega=Dq^2\taum$. (d) Simulated normalized transport response (symbols) compared with the analytical transfer function (solid line), demonstrating quantitative agreement over the investigated transport-memory range.}
\label{fig:2}
\end{figure}

The single-exponential kernel is used below as an analytically transparent reference; other memory kernels preserve the general relation in Eq.~\eqref{eq:6}.

\subsection{Anisotropic diffusion}
\label{sec:anisotropic}

The analytical forward model developed above extends naturally from isotropic to anisotropic diffusion. For homogeneous anisotropic diffusion, stochastic transport is characterized by the symmetric diffusion tensor $\bD$~\cite{Basser}. The Gaussian displacement propagator (Supplemental Material Sec.~S2) has characteristic function
\begin{equation*}
\Phi(\bq,\tau) = \exp(-\bq^{T}\bD\bq\,\tau),
\end{equation*}
so that $R(\bq)=H(\bq^{T}\bD\bq)$. Anisotropy therefore modifies only the transport variable, not the transfer function $H$; isotropic and anisotropic Brownian diffusion are thus described by the common forward relation
\begin{equation}
R(\bq) = H[T(\bq)], \qquad T(\bq) = \bq^{T}\bD\bq,
\label{eq:8}
\end{equation}
which reduces identically to $T=Dq^2$ for isotropic diffusion, $\bD=D\mathbf{I}$. For the single-exponential reference model, the corresponding directional transport-memory parameter is $\Omega(\bq)=T(\bq)\taum$.

\section{Stochastic Diffusion Inversion in OMT Imaging}
\label{sec:inversion}

The forward model established in the preceding section predicts how stochastic diffusion modifies the spatial modulation of the optical-memory emission. OMT imaging addresses the corresponding inverse problem: recovering the underlying diffusion parameters from this diffusion-dependent optical response. Experimentally, the modulation component at the imposed illumination wavevector is extracted from the recorded emission image following the acquisition and demodulation procedure established previously for OMT imaging~\cite{OMTflow}. To isolate the diffusion-dependent contribution, the measured modulation coefficient is normalized by a zero-diffusion reference acquired under otherwise identical conditions,
\begin{equation}
R(\bq) = \frac{S_{\bq}}{S_{\bq}^{0}},
\label{eq:9}
\end{equation}
where $S_{\bq}$ and $S_{\bq}^{0}$ are the modulation coefficients measured with and without diffusion, respectively. We first consider a spatially uniform scalar diffusion coefficient before extending the inversion to spatially varying scalar and tensor diffusion. The calibration conditions, uniqueness of the inverse response, and numerical scalar and tensor reconstruction procedures are detailed in Sec.~S3 of the Supplemental Material.

\subsection{Constant diffusion coefficient inversion}
\label{sec:constant}

Once the measured image has been reduced to the normalized response $R(\bq)$, reconstruction of a spatially uniform diffusion coefficient becomes a one-dimensional inversion problem. For the single-exponential memory kernel, the normalized diffusion response is monotonic and can be inverted analytically,
\begin{equation}
\widehat\Omega = \frac{1}{R}-1, \qquad \widehat D = \frac{\widehat\Omega}{q^2\taum},
\label{eq:10}
\end{equation}
where $\widehat D$ denotes the reconstructed diffusion coefficient. More generally, an analytical form is not required: writing the calibrated inverse-response map as $\mathcal C$, so that $\widehat\Omega=\mathcal C(R)$, the same inversion can be performed using an experimentally calibrated response curve, provided the response is one-to-one over the operating range.

\begin{figure*}[t]
\centering
\includegraphics[width=0.85\textwidth]{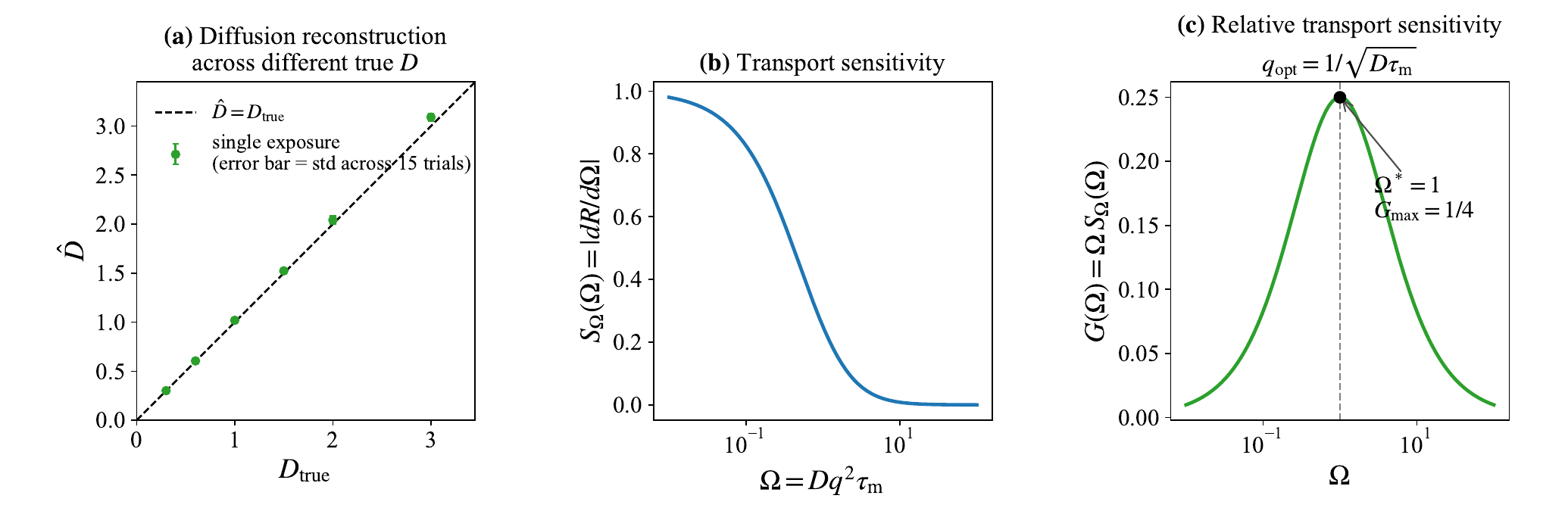}
\caption{Diffusion inversion, transport sensitivity, and measurement design. (a) Diffusion coefficients reconstructed from the calibrated transfer-function inversion versus ground truth, demonstrating accurate inversion of the stochastic OMT response. (b) Absolute transport sensitivity $S_\Omega(\Omega)=|dR/d\Omega|$, showing the monotonic decrease in response sensitivity with increasing transport-memory parameter. (c) Relative transport sensitivity $G(\Omega)=\Omega S_\Omega(\Omega)$, which sets the response-sensitivity factor entering the Fisher information for relative diffusion estimation and exhibits an optimal operating point at $\Omega=1$. This optimum provides a design criterion for selecting the structured-illumination spatial frequency.}
\label{fig:3}
\end{figure*}

Figure~\ref{fig:3}(a) demonstrates this inversion for isotropic Brownian diffusion. Diffusion coefficients reconstructed from independently generated Monte Carlo responses follow the identity line over the investigated range, confirming that the stochastic OMT response is not only predicted by the forward model but can also be quantitatively inverted to recover the diffusion coefficient.

The ability to invert the response raises a closely related question: how strongly does the measured response change for a given change in diffusion? We define the transport sensitivity
\begin{equation}
S_\Omega(\Omega) \equiv \left|\frac{dR}{d\Omega}\right|,
\label{eq:11}
\end{equation}
which for the single-exponential kernel, using Eq.~\eqref{eq:7}, evaluates to $S_\Omega(\Omega)=1/(1+\Omega)^2$. As shown in Fig.~\ref{fig:3}(b), $S_\Omega$ decreases monotonically with increasing $\Omega$: stronger diffusion-induced averaging progressively flattens the optical response, so that a given additional change in $\Omega$ produces a smaller measurable response variation.

The absolute response slope alone, however, does not determine how effectively the diffusion coefficient itself is encoded. Because $\Omega=Dq^2\taum$, a fractional change in $D$ produces the same fractional change in $\Omega$. We therefore define the relative transport sensitivity
\begin{equation}
G(\Omega) \equiv \Omega S_\Omega(\Omega),
\label{eq:12}
\end{equation}
which for the single-exponential response is $G(\Omega)=\Omega/(1+\Omega)^2$. Unlike $S_\Omega$, which decreases monotonically, $G(\Omega)$ vanishes in both limiting regimes: for $\Omega\to0$, the response slope is largest but the transport-memory coupling itself is weak, while for $\Omega\to\infty$, strong averaging flattens the response and suppresses further measurable change. The competition between these limits produces the maximum shown in Fig.~\ref{fig:3}(c),
\begin{equation}
\Omega^{*}=1,\quad G_{\max}=\frac14,\quad\text{giving}\quad q_{\opt}=\frac{1}{\sqrt{D\taum}}.
\label{eq:13}
\end{equation}
Thus, the illumination spatial frequency can be selected to place the expected diffusion coefficient near the regime of maximal relative response sensitivity. The statistical precision additionally depends on the response noise, as quantified by the Fisher-information analysis developed below.

\subsection{Spatial scalar diffusion inversion}
\label{sec:spatial}

The constant-diffusion inversion developed above extends to spatially heterogeneous diffusion when the field is locally uniform over the spatiotemporal range sampled by the OMT measurement. Each image region is then described by a local diffusion coefficient $D(\br)$. The local response is inverted using the same calibrated relation as the spatially uniform case,
\begin{equation}
\widehat\Omega(\br) = \mathcal C[R(\br)], \qquad \widehat D(\br) = \frac{\widehat\Omega(\br)}{q^2\taum},
\label{eq:14}
\end{equation}
where $\mathcal C$ denotes the calibrated inverse-response map introduced above, evaluated at the locally measured response.

To test this spatial extension, Monte Carlo simulations were performed for a prescribed heterogeneous diffusion landscape in which $D$ varies continuously along the $x$ direction. Figure~\ref{fig:4}(a) presents the resulting optical-memory emission image; regions of lower diffusivity retain stronger modulation, while increasing diffusivity produces progressively weaker modulation. As shown in Fig.~\ref{fig:4}(b), the reconstructed $D(x)$ closely follows the prescribed profile, demonstrating quantitative recovery of the heterogeneous scalar diffusion landscape.

\subsection{Tensor diffusion inversion}
\label{sec:tensor}

When diffusion is anisotropic, the local process is described by a symmetric diffusion tensor $\bD(\br)$, whose unequal principal diffusivities characterize the directional dependence of transport. An illumination wavevector $\bq_j$ probes the direction-dependent transport-memory parameter, which in two dimensions is
\begin{equation}
\begin{split}
\Omega_j(\br) = \taum\bq_j^{T}\bD(\br)\bq_j = \taum\big[&q_{jx}^2D_{xx}(\br)+2q_{jx}q_{jy}D_{xy}(\br)\\
&{}+q_{jy}^2D_{yy}(\br)\big].
\end{split}
\label{eq:15}
\end{equation}
Inversion of the calibrated OMT response for each illumination direction yields one local quadratic projection $\widehat\Omega_j(\br)$; measurements along $M$ directions combine, via the illumination matrix
\begin{equation*}
\bQ = \begin{pmatrix}
q_{1x}^2 & 2q_{1x}q_{1y} & q_{1y}^2 \\
q_{2x}^2 & 2q_{2x}q_{2y} & q_{2y}^2 \\
\vdots & \vdots & \vdots \\
q_{Mx}^2 & 2q_{Mx}q_{My} & q_{My}^2
\end{pmatrix},
\end{equation*}
to reconstruct the tensor components,
\begin{equation}
\widehat{\bd}(\br) = \frac{1}{\taum}\bQ^{+}\widehat{\bOmega}(\br),
\label{eq:16}
\end{equation}
where $\widehat{\bd}(\br)=[\widehat D_{xx}(\br),\widehat D_{xy}(\br),\widehat D_{yy}(\br)]^{T}$, $\widehat{\bOmega}(\br)=[\widehat\Omega_1(\br),\widehat\Omega_2(\br),\ldots,\widehat\Omega_M(\br)]^{T}$, and $\bQ^{+}$ denotes the Moore--Penrose pseudoinverse of $\bQ$. At least three independent illumination directions are required to determine the three independent components of a symmetric two-dimensional diffusion tensor, while additional directions provide an overdetermined reconstruction.

We tested this inversion using Monte Carlo simulations of a prescribed heterogeneous anisotropic diffusion field. Brownian trajectories were propagated through the spatially varying diffusion-tensor field, and separate optical-memory emission images were generated for each illumination direction (see Sec.~S6 of the Supplemental Material). For each directional image, the local modulation was extracted and normalized to the corresponding zero-diffusion reference to obtain the directional diffusion response $R_j(\br)$, which was then inverted through the calibrated response relation to yield the local transport-memory projection $\widehat\Omega_j(\br)$. The resulting set of directional projections was combined through Eq.~\eqref{eq:16} to reconstruct the local diffusion tensor $\widehat{\bD}(\br)$.

\begin{figure*}[t]
\centering
\includegraphics[width=0.95\textwidth]{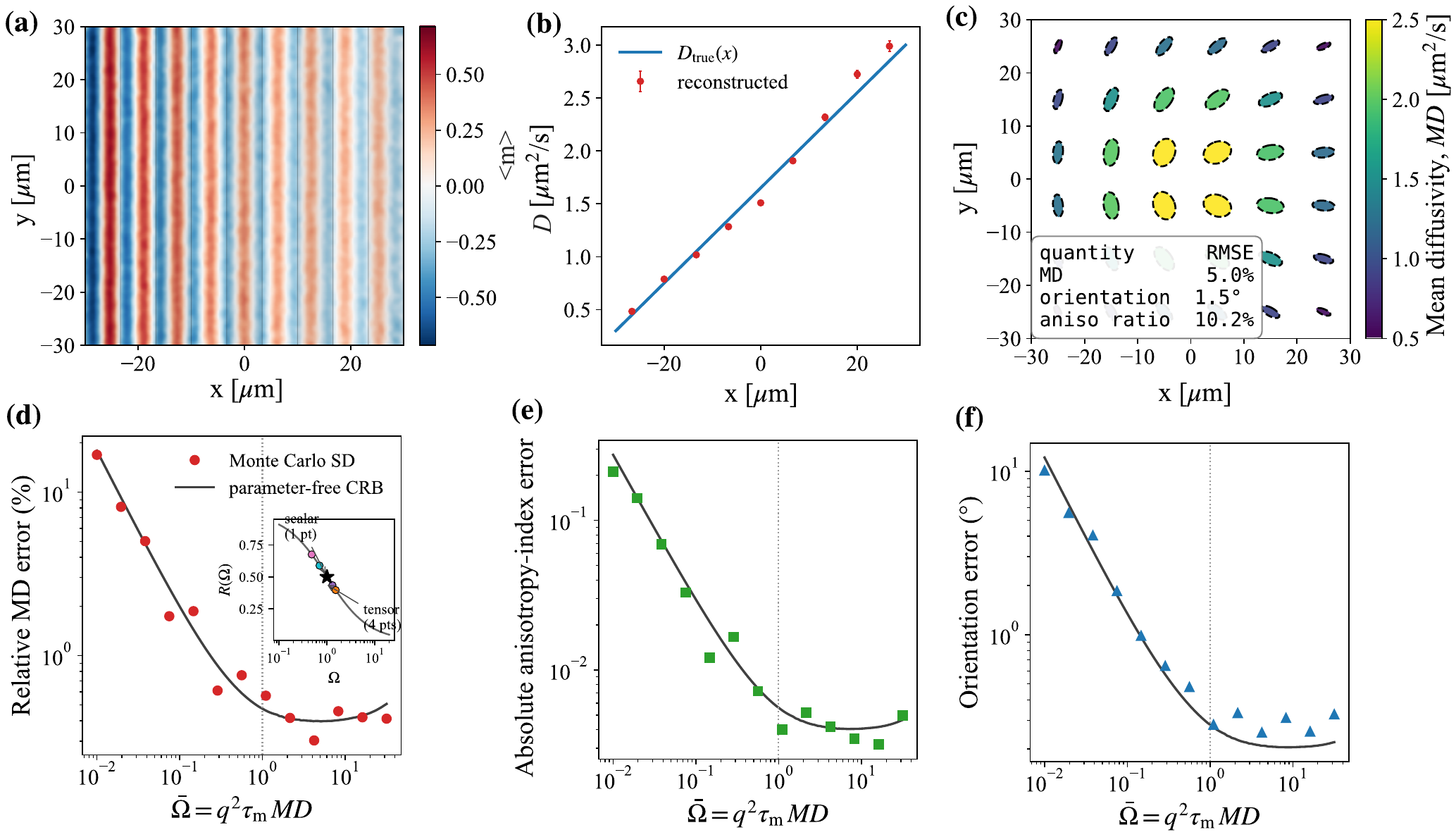}
\caption{Spatial and tensor diffusion inversion and precision limits. (a) Simulated optical-memory emission image for a heterogeneous diffusion landscape under structured illumination. Spatial variations in fringe contrast reflect the underlying diffusion coefficient through transport-induced averaging. (b) Reconstructed scalar diffusion profile obtained by local transfer-function inversion, compared with the prescribed ground truth. (c) Tensor diffusion imaging from multiple structured-illumination directions. The reconstructed tensor field captures the prescribed anisotropic transport, with root-mean-square errors for the mean diffusivity ($MD$), principal orientation, and anisotropy ratio summarized in the panel. (d) Monte Carlo standard deviation of the reconstructed mean diffusivity compared with the first-principles, parameter-free Cram\'er--Rao bound as a function of the mean transport-memory parameter $\bar\Omega$. The analytical prediction uses no fitted scaling factor or simulation-derived noise calibration. The inset illustrates the response operating points sampled in scalar and four-direction tensor measurements. (e,f) Corresponding comparisons for the diffusion anisotropy $A$ and principal-axis orientation, respectively.}
\label{fig:4}
\end{figure*}

Figure~\ref{fig:4}(c) shows the reconstructed tensor field from multiple directional acquisitions, capturing the prescribed spatial variations in mean diffusivity, anisotropy ratio, and principal-axis orientation with low root-mean-square errors across all three descriptors. The attainable precision of these reconstructions is analyzed in the following section.

\section{Precision Limits of Stochastic Diffusion Imaging}
\label{sec:precision}

The inversion framework developed in the preceding section establishes how scalar and tensor diffusion parameters can be reconstructed from the measured OMT response. The attainable precision of these estimates, however, depends jointly on the response sensitivity and measurement noise. For a Gaussian-distributed response estimate with mean $R(\Omega)$ and variance $\sigma_R^2(\Omega)$, the full Fisher information~\cite{Kay} contains contributions from the transport dependence of both the mean and the variance (Supplemental Material Sec.~S5). Over the complete operating range considered here, the variance-derivative contribution is below $0.01\%$ of the mean-response contribution. The Fisher information associated with the transport-memory parameter $\Omega$ can therefore be written, to excellent accuracy, as
\begin{equation}
\mathcal I(\Omega) \simeq \frac{1}{\sigma_R^2(\Omega)}\left[\frac{dR(\Omega)}{d\Omega}\right]^2 = \frac{S_\Omega^2(\Omega)}{\sigma_R^2(\Omega)},
\label{eq:17}
\end{equation}
where $S_\Omega(\Omega)$ is the transport sensitivity introduced in Eq.~\eqref{eq:11}. Using this leading-order expression, the corresponding Cram\'er--Rao lower bound is
\begin{equation}
\operatorname{Var}(\widehat\Omega) \geq \frac{1}{\mathcal I(\Omega)} = \frac{\sigma_R^2}{S_\Omega^2(\Omega)},
\label{eq:18}
\end{equation}
where $\operatorname{Var}(\widehat\Omega)$ denotes the variance of an unbiased estimator $\widehat\Omega$.

For isotropic Brownian diffusion, $\Omega=Dq^2\taum$, and propagating the Fisher information through this parameter transformation (Supplemental Material Sec.~S5) gives the relative Cram\'er--Rao bound
\begin{equation}
\frac{\operatorname{Var}(\widehat D)}{D^2} \geq \frac{\sigma_R^2(\Omega)}{\Omega^2 S_\Omega^2(\Omega)} = \frac{\sigma_R^2(\Omega)}{G^2(\Omega)},
\label{eq:19}
\end{equation}
where $G(\Omega)=\Omega S_\Omega(\Omega)$ is the relative transport sensitivity introduced in Eq.~\eqref{eq:12}.

For anisotropic diffusion, measurements acquired along multiple illumination directions jointly constrain the tensor parameters. We parameterize the two-dimensional diffusion tensor by
\begin{equation*}
\bmeta = (MD,A,\theta),
\end{equation*}
where $MD$ is the mean diffusivity, $A=(D_\parallel-D_\perp)/(D_\parallel+D_\perp)$ is the normalized diffusion anisotropy, and $\theta$ is the principal-axis orientation. For an illumination wavevector of fixed magnitude $q$ and direction $\phi_j$, the corresponding transport-memory parameter is
\begin{equation*}
\Omega_j = \bar\Omega\left[1+A\cos2(\phi_j-\theta)\right], \qquad \bar\Omega = MD\,q^2\taum.
\end{equation*}
Each illumination direction therefore samples the calibrated OMT response at a different tensor-dependent operating point. The inset of Fig.~\ref{fig:4}(d) illustrates this distinction: a scalar diffusion measurement corresponds to a single operating point on the response curve, whereas the tensor measurement considered here samples four direction-dependent operating points.

For statistically independent directional response estimates, and neglecting the variance-derivative contribution justified above, the Fisher information generalizes to the Fisher-information matrix~\cite{Kay}
\begin{equation}
F_{ab} = \sum_j \frac{S_\Omega^2(\Omega_j)}{\sigma_{R,j}^2}\frac{\partial\Omega_j}{\partial\eta_a}\frac{\partial\Omega_j}{\partial\eta_b},
\label{eq:20}
\end{equation}
with the corresponding Cram\'er--Rao bound
\begin{equation}
\operatorname{Var}(\widehat\eta_a) \geq (\mathbf F^{-1})_{aa}.
\label{eq:21}
\end{equation}
The derivatives entering the Fisher matrix, together with the corresponding generalization for parameter-dependent response variances, are given in Sec.~S5 of the Supplemental Material. These expressions show that tensor-reconstruction precision is determined jointly by the OMT response sensitivity, measurement noise, and directional sampling geometry.

To obtain an absolute precision prediction, the directional response variances $\sigma_{R,j}^2$ must also be specified. For the finite-particle and finite-exposure conditions considered here, these variances are calculated analytically from the finite-sampling statistics of the memory-filtered optical response (see Sec.~S4 of the Supplemental Material), requiring neither a fitted scaling factor nor noise calibration from the Monte Carlo simulations; the Cram\'er--Rao bounds below therefore provide first-principles, parameter-free precision predictions.

Figures~\ref{fig:4}(d)--\ref{fig:4}(f) provide an independent test of these parameter-free precision predictions as functions of the mean transport-memory parameter $\bar\Omega$. Figure~\ref{fig:4}(d) compares the Monte Carlo standard deviation of the reconstructed mean diffusivity with the first-principles Cram\'er--Rao bound, while Figs.~\ref{fig:4}(e) and \ref{fig:4}(f) show the corresponding comparisons for the diffusion anisotropy $A$ and principal-axis orientation, respectively. Across the investigated transport-memory range, the Monte Carlo precision closely follows the analytical bounds for all three reconstructed quantities, without introducing any fitted scaling factor or simulation-derived noise calibration. Additional validation of the analytical response-variance model and numerical bias diagnostics are provided in Sec.~S7 of the Supplemental Material.

The orientation precision additionally reflects a fundamental identifiability limit of anisotropic diffusion. As $A\to0$,
\begin{equation*}
\frac{\partial\Omega_j}{\partial\theta} = 2A\bar\Omega\sin2(\phi_j-\theta) \to 0,
\end{equation*}
so that the Fisher information associated with $\theta$ vanishes, causing its Cram\'er--Rao bound to diverge (see Supplemental Material Sec.~S5 for the complete Fisher-matrix argument). The loss of orientation precision near isotropic diffusion occurs because the principal direction becomes physically ill-defined as the anisotropy vanishes.

In experimental applications, the same Fisher-information framework can be used with independently characterized response-noise statistics, obtained either from an analytical acquisition model or from repeated calibration measurements.

\section*{Discussion}

The present results establish optical-memory transport (OMT) imaging as a distinct route to diffusion measurement. Conventional diffusion-imaging approaches commonly infer transport through an externally resolved temporal coordinate, such as a recovery time, frame delay, transient decay, or scanning interval. OMT instead uses the intrinsic causal dynamics of the tracer to weight past transport histories, while structured illumination converts this memory-weighted history into a spatially encoded emission response. Diffusion can therefore be inferred from a steady-state modulation change without explicitly resolving the temporal evolution of the sample. This architecture reduces the reliance on time-resolved acquisition characteristic of recovery-, frame-delay-, and scanning-based measurements. Its practical advantages under specific experimental conditions, however, will depend on factors such as photon statistics, tracer photophysics, acquisition constraints, and spatial resolution, and require dedicated experimental validation.

The transport-history integral provides the stochastic generalization underlying this mechanism. In deterministic OMT, a tracer detected at a given position is associated with a unique upstream trajectory, whereas stochastic transport replaces that trajectory by a conditional distribution over all compatible upstream histories. The deterministic formulation is recovered when the conditional propagator collapses to a delta function. Deterministic flow and stochastic diffusion therefore emerge as two limits of the same transport-history formulation.

Structured illumination provides the connection between this stochastic transport description and an experimentally accessible optical response. A spatial Fourier mode with wavevector $\bq$ probes the characteristic function of the conditional displacement distribution at the same wavevector. For homogeneous Brownian diffusion, this characteristic function decays as $\exp(-Dq^2\tau)$, reducing the OMT response to the Laplace transform of the optical-memory kernel evaluated at the diffusion transport variable $T=Dq^2$. The illumination wavevector therefore sets the spatial scale of the diffusion measurement, while the optical-memory time sets the temporal range over which transport history contributes. For the single-exponential reference model, this coupling is governed by the dimensionless transport-memory parameter $\Omega=Dq^2\taum$. The resulting dependence of measurement sensitivity on $\Omega$ also provides a direct experimental-design principle: the illumination spatial frequency can be selected to place the expected diffusion range near the regime of maximal relative response sensitivity.

The Fisher-information analysis further places these inversions within a common statistical framework, relating reconstruction precision to the optical-response sensitivity, finite-sampling noise, and directional sampling geometry. Because the response fluctuations can themselves be predicted analytically, the resulting Cram\'er--Rao bounds provide parameter-free predictions of scalar and tensor reconstruction precision without empirical calibration to the Monte Carlo results.

Practical quantitative imaging requires the OMT response to be calibrated under the same photophysical operating conditions used for measurement. This is particularly important for nonlinear optical-memory probes, whose effective response can depend on excitation conditions and population dynamics. Spatially resolved reconstruction also relies on local transport uniformity over the region used to estimate the modulation response, introducing a trade-off between spatial localization and statistical precision. Within these constraints, however, the inversion requires only a calibrated relation between the measured optical response and the relevant transport-memory parameter; it does not require the optical-memory dynamics to follow the single-exponential model used here as an analytically transparent reference. This separation between transport dynamics and tracer photophysics provides a modular interface through which other optical-memory probes can be incorporated into the same transport framework.

Although the present work focuses on stationary Brownian diffusion and its tensor generalization, the transport-history formulation is not restricted to Gaussian diffusion. More general stochastic processes, including confined or compartmentalized and anomalous diffusion, can in principle be incorporated through their conditional displacement statistics and the corresponding characteristic functions. Measurements at different illumination wavevectors provide access to transport over different spatial scales. Similar in spirit to spot-variation and variable-length-scale FCS, in which diffusion is probed over different observation areas~\cite{Wawrezinieck2005,Humpolickova2006,Veerapathiran2018}, and to X-ray photon correlation spectroscopy, in which dynamics are examined as a function of scattering wavevector ~\cite{Perakis2025}, measurements at multiple illumination wavevectors may reveal departures from the Brownian response or help distinguish between candidate transport models when sufficient independent measurements are available. Where an explicit transport model is unavailable or impractical, the corresponding transport-response relation may instead be established empirically using reference measurements or numerical benchmarks under known transport conditions.

Together with the deterministic-flow formulation of OMT, the present stochastic extension establishes finite optical memory as a common physical basis for transport imaging across deterministic and stochastic regimes. By encoding transport history through intrinsic tracer dynamics rather than an externally resolved time coordinate, OMT provides a unified framework in which flow, scalar diffusion, and anisotropic diffusion can be described, inverted, and quantitatively assessed within the same measurement principle.

\begin{acknowledgments}
\emph{Acknowledgments}--- The authors used OpenAI ChatGPT (GPT-5.6 Sol)
and Anthropic Claude (Claude Sonnet 5, accessed through Claude Code)
during manuscript preparation as auxiliary tools for consistency
checking, cross-checking mathematical derivations and numerical
results, and refining scientific exposition. All AI-assisted
suggestions relevant to the scientific content were critically
evaluated by the authors and independently verified, where
appropriate, against analytical derivations, numerical calculations,
and source code. The authors take full responsibility for the
scientific content and conclusions. H.L. and J.W. acknowledge
financial support from the ÅForsk Foundation (23-322), the Carl
Trygger Foundation (23-2635), and the Swedish Research Council
(VR 2021-04556, 2025-05609).
\end{acknowledgments}

\emph{Data availability}--- The data and numerical results that support the findings of this study are available from the corresponding author upon reasonable request.

\emph{Conflict of interest}--- H.L. and J.W. are inventors on a pending patent application covering aspects of stochastic transport measurement using finite-memory optical reporters and spatially structured excitation described in this work.

\bibliographystyle{apsrev4-2}
\bibliography{main-20260922}

\end{document}